\documentclass{article}
\usepackage{amssymb}

\usepackage{graphicx}
\usepackage{amsmath}

\input{tcilatex}

\begin{document}

\title{\textbf{POSSIBLE}\ \textbf{SIGNATURE OF LOW SCALE GRAVITY IN ULTRA HIGH
ENERGY COSMIC RAYS}}
\author{\textbf{R.V.Konoplich}$^{1}$\textbf{, S.G.Rubin}$^{2,3}$ \\
\\
$^{1}$New York University, New York, USA\\
$^{2}$Moscow Engineering Physics Institute, Moscow,Russia\\
$^{3}$Center for Cosmoparticle Physics ''Cosmion'', Moscow, Russia}
\maketitle

\begin{abstract}
We show that the existence of low scale gravity at TeV scale could lead to a
direct production of photons with energy above 10$^{22}$ eV due to
annihilation of ultra high energy neutrinos on relic massive neutrinos of
the galactic halo. Air showers initialized in the terrestrial atmosphere by
these ultra energetic photons could be collected in near future by the new
generation of cosmic rays experiments.
\end{abstract}

\ \ \ \ Recently it was proposed \cite{1} that the space is 4+n dimensional,
with the Standard Model particles living \ \ \ on a brane. While the weakly,
electromagnetically, and strongly interacting particles are confined to the
brane in 4 dimensions, gravity can propagate also in extra n dimensions.
This approach allows to save the gauge hierarchy problem by introducing a
single fundamental mass scale (string scale) $M_{s}$ of the order of TeV.
The usual Planck scale $M_{Pl}=1/\sqrt{G_{N}}\simeq 1.22\cdot 10^{19}$GeV is
related to the new mass scale $M_{s}$ by Gauss's law:

\begin{equation}
M_{Pl}^{2}\sim R^{n}M_{s}^{n+2}  \label{1}
\end{equation}

where $G_{N}$ is the Newton constant, $R$ is the size of extra dimensions.
It follows from (\ref{1}) that

\begin{equation}
R\sim 2\cdot 10^{-17}(\frac{TeV}{M_{s}})(\frac{M_{Pl}}{M_{s}})^{2/n},cm
\label{2}
\end{equation}

gives at $n=1$ too large value, which is clearly excluded by present
gravitation experiments. On the other hand $n\geq 2$ gives the value $%
R\lesssim 0.25$ cm, which is below the present experimental limit $\sim $1
cm but can be tested for the case $n=2$ in gravitational experiments in near
future.

It can be shown that the graviton including its excitations in the extra
dimensions, so-called Kaluza-Klein (KK) graviton emission, interacts with
the Standard Model particles on the brane with an effective amplitude $\sim
M_{s}^{-1}$ instead of $M_{Pl}^{-1}$. Indeed, the graviton coupling to the
Standard Model particle $\sim M_{Pl}^{-1}$, the rate \cite{2} of the
graviton interaction $r\sim (M_{Pl}^{-1})^{2}N$, where $N$ is a multiplicity
of KK-states. Since this factor is $\sim $ ($\sqrt{S}R)^{n}$, where $\sqrt{S}
$ is the c.m. energy, then substituting $R$ from (\ref{2}) we get $r\sim
M_{s}^{-2}$. Thus the graviton interaction becomes comparable in strength
with weak interaction at TeV scale.

This leads to the varieties of new signatures in particle physics,
astrophysics and cosmology (see e.g. \cite{2,3,4,5,6} ) which have already
been tested in experiments or can be tested in near future.

In this article we consider the possible signature of the low scale gravity
in ultra high energy cosmic rays.

The detection \cite{7,8} of cosmic rays with energy above
Greisen-Zatsepin-Kuzmin (GZK) cut-off of $\sim 5\cdot 10^{19}eV$ presents a
serious problem for interpretation. The origin of GZK cut-off \cite{9} is
due to resonant photoproduction of pions by protons on cosmic microwave
background radiation which leads to a significant degradation of proton
energy (about 20\% for 6 Mpc) during its propagation in the Universe. Of
course, proton energy does not change by many orders of magnitude if high
energy protons come from the distances \TEXTsymbol{<} 50 - 100 Mpc. However,
no nearby sources like active galactic nuclei have been found up to now in
the arrival direction.

It is difficult also to relate the observed ultra high energy events (Fig.1 
\cite{8}) with the other particles. For example in the case of ultra high
energy photons due to interaction with cosmic background radiation ($\gamma
+\gamma ^{\ast }\longrightarrow e^{+}+e^{-}$) the photon free mean path
should be significantly less than 100 Mpc. A scenario based on direct cosmic
neutrinos able to reach the Earth from cosmological distances can not
reproduce the observed signatures of ultra high energy air showers occurred
high in the atmosphere.

Different possibilities were considered (see e.g. \cite{10} and references
therein) in order to solve this puzzle. In particular it was proposed \cite
{11,12,13} that ultra high energy neutrinos reaching the Earth from
cosmological distances interact with a halo of relic light neutrinos in the
Galaxy, producing due to Z, W$^{\pm }$ boson exchange secondaries inside the
galactic halo. Photons from $\pi ^{0}$ decays and nucleons can easily
propagate to the Earth and be the source of the observed ultra high energy
air showers. Critical elements of models \cite{11,12,13} are: the existence
of neutrino mass in the range 0.1-10 eV and significant clustering of relic
neutrinos in the halo up to 10$^{5}$n$_{\nu }$, where n$_{\nu }$ is the
cosmological neutrino number density (n$_{\nu }$ $\sim 100cm^{-3}$). Also
the existence of ultra high energy (\TEXTsymbol{>} 10$^{21}$-10$^{23}$ eV)
neutrino flux is necessary in order to produce multiple secondaries with
energies above GZK cut-off.

However if the graviton interaction comparable in strength with weak
interaction at TeV scale exists then photons can be produced directly in a
reaction 
\begin{equation}
\nu +\overset{-}{\nu }\longrightarrow g\longrightarrow \gamma +\gamma
\label{3}
\end{equation}

due to virtual graviton exchange (Fig.2). In Standard Model process (\ref{3}%
) occurs via loop diagram and therefore is severely suppressed.

At high energies the cross section for the process (\ref{3}) can be obtained
immediately from that for the process $e^{+}e^{-}\longrightarrow \gamma
\gamma $ including graviton exchange (see for example \cite{3}) by
substituting e = 0. Then 
\begin{equation}
\frac{d\sigma }{dz}=\frac{\pi }{16}\frac{S^{3}}{M_{s}^{8}}F^{2}(1-z^{4})
\label{4}
\end{equation}

where $\sqrt{S}$ is c.m.s. energy, z = $\left| \cos \theta \right| $ is the
polar angle of the outgoing photon. The factor F depends on the number of
extra dimensions: 
\begin{equation*}
F= 
\begin{array}{c}
\log (M_{s}^{2}/S)\underset{}{} \\ 
2/(n-2)\overset{}{}
\end{array}
, 
\begin{array}{c}
n=2\underset{}{} \\ 
n>2\overset{}{}
\end{array}
,
\end{equation*}

at $\sqrt{S}<<M_{s}$. In Eq.(\ref{4}) it is also taken into account that
primary beam of neutrinos is polarized.

Integrating (\ref{4}) over the polar angle and including a symmetry factor
for two $\gamma $ we get 
\begin{equation}
\sigma =\frac{\pi }{20}\frac{S^{3}}{M_{s}^{8}}F^{2}\approx 7\cdot
10^{-35}F^{2}\QOVERD( ) {\sqrt{S}}{TeV}^{6}\QOVERD( ) {TeV}{M_{s}}^{8}cm^{2}.
\label{5}
\end{equation}

On can see from (\ref{5}) that at TeV energies the rate of the reaction (\ref
{3}) is comparable with the rate of weak processes \cite{11}.

Assuming $M_{s}\sim $ $\sqrt{S}\sim $ TeV we find for example for $n=3$ the
following probability for the interaction of ultra high energy neutrinos
inside the galactic halo: $P\approx \sigma n_{G}L_{G}\sim 10^{-3}$, where $%
L_{G}\sim 100$ Kpc is the size of the galactic neutrino halo, $n_{G}\sim
10^{5}n_{\nu }$ is the neutrino number density in the galactic halo. This
probability is significantly greater than the probability of ultra high
energy neutrino interaction in terrestrial atmosphere \cite{14}.

Let us note that nearby galaxies also can be sources of additional ultra
high energy photons due to neutrino interaction with relic neutrinos of
galactic halos \cite{12,13}.

TeV range in c.m.s. corresponds to the energy of extragalactic neutrino flux 
$E\approx 10^{22}-10^{23}$ eV since 
\begin{equation}
E\approx \frac{S}{2m}\approx 5\cdot 10^{22}\QOVERD( ) {\sqrt{S}%
}{TeV}^{2}\QOVERD( ) {10eV}{m}eV  \label{6}
\end{equation}

where m is neutrino mass.

Photon distribution in reaction (\ref{3}) in laboratory system is given by 
\begin{equation}
\frac{d\sigma }{d(\omega /E)}=8\pi F^{2}\frac{m^{3}E^{3}}{M_{s}^{8}}\frac{%
\omega }{E}(1-\frac{\omega }{E})[(1-\frac{\omega }{E})^{2}+(\frac{\omega }{E}%
)^{2}]  \label{7}
\end{equation}

where $\omega >>m$ is photon energy. This distribution is shown in Fig.3. It
follows from (\ref{7}) that photons are produced in the reaction (\ref{3})
mainly within the energy range $0.2E\lesssim \omega \lesssim 0.8E$ with an
average energy $\approx E/2$.

Therefore existence of low scale gravity at TeV scale or above could lead to
the direct production of photons with energy $\omega >10^{22}$ eV (at these
energies the mean interaction length for pair production for photons in the
radio background is $\approx 1-10Mpc$ \cite{15}). Such photons can be hardly
produced in standard weak interaction processes because in last ones photons
appear as a result of cascade processes significantly reducing photon energy
in comparison with the initial neutrino energy. For example, as it was shown
in \cite{11} final energy of photons produced due to cascade processes can
be by 10-100 times less than the energy of the initial neutrino flux.

Of course photons with the energy $\sim 10^{23}$ eV could be produced in
cascade processes induced by neutrinos of the energy \TEXTsymbol{>} $%
10^{24}-10^{25}$ eV but from the observations of cosmic rays we know that
cosmic ray fluxes decrease with the energy as $E^{-3}$, and therefore the
probability of such events is significantly suppressed.

Fluxes of ultra high energy cosmic rays at the Earth are very small $\Phi
\sim 0.03km^{-2}sr^{-1}yr^{-1}$. Until now only about 60 events were
collected with energies above GZK cut-off. However in near future improved
Fly's Eye (7000 $km^{2}sr$) \cite{8} will allow to detect about 20
events/yr. It seems possible that such detector could collect rare ultra
energetic photons ($\omega >10^{22}eV$). The detection of such events could
be an indication that these ultra high energy photons were produced in $\nu 
\overset{-}{\nu }$ annihilation in the galactic halo due to effects of low
scale gravity at TeV scale.

Authors thank D.Fargion for interesting discussions on ultra high energy
cosmic rays. One of us (RVK) is grateful to Physics Department of New York
University for warm hospitality.

\end{document}